\documentclass[twocolumn,pra,superscriptaddress]{revtex4-1}
\usepackage[paperwidth=210mm,paperheight=297mm,centering,hmargin=2.0cm,vmargin=2.6cm]{geometry}
\usepackage[latin1]{inputenc}
\usepackage{amsmath}
\usepackage{amsfonts}
\usepackage{graphics}
\usepackage{epsfig}
\usepackage{dcolumn}

\newcommand{\mcE}{\hat{\mathcal{E}}}

\newcommand{\psiscat}{\psi_{\text{scat}}}

\newcommand{\beqa}{\begin{eqnarray}}
\newcommand{\eeqa}{\end{eqnarray}}

\newcommand{\ee}{\mathrm{e}}
\newcommand{\ii}{\mathrm{i}}

\newcommand{\bra}[1]{\langle #1 |}
\newcommand{\ket}[1]{|#1\rangle}
\newcommand{\braket}[2]{\langle #1 | #2 \rangle}

\newcommand{\beq}[1]{\begin{equation} \eqlab{#1}}
\newcommand{\eeq}{\end{equation}}
\newcommand{\bsub}{\begin{subequations}}
\newcommand{\esub}{\end{subequations}}
\newcommand{\nn}{\nonumber}

\newcommand{\eqlab}[1]{\label{eq:#1}}
\renewcommand{\eqref}[1]{Eq.~(\ref{eq:#1})}
\newcommand{\eqsref}[2]{Eqs.~(\ref{eq:#1}) and~(\ref{eq:#2})}
\newcommand{\figref}[1]{Fig.~\ref{fig:#1}}
\newcommand{\figsref}[2]{Figs.~\ref{fig:#1} and~\ref{fig:#2}}
\newcommand{\figlab}[1]{\label{fig:#1}}

\newcommand{\secref}[1]{Section~\ref{sec:#1}}
\newcommand{\secsref}[2]{Sections~\ref{sec:#1} and~\ref{sec:#2}}
\newcommand{\seclab}[1]{\label{sec:#1}}

\newcommand{\onefigadv}[5]{
\begin{figure}[#5]
\centering
\includegraphics[width=#4\textwidth]{#1}
\\[-7pt]
\caption{#2}\figlab{#3}
\end{figure}
}

\begin{document}

\title{Limitations of two-level emitters as non-linearities in two-photon \\ controlled phase gates}
\author{Anders Nysteen}
\affiliation{DTU Fotonik, Department of Photonics Engineering, Technical University of Denmark, Building 343, 2800 Kgs. Lyngby, Denmark}
\author{Dara P. S. McCutcheon}
\affiliation{Quantum Engineering Technology Labs, H. H. Wills Physics Laboratory and Department of Electrical and Electronic Engineering, 
University of Bristol, Merchant Venturers Building, Woodland Road, Bristol BS8 1FD, UK}
\author{Mikkel Heuck}
\affiliation{Department of Electrical Engineering and Computer Science, Massachusetts Institute of Technology,
Cambridge, Massachusetts 02139, USA}
\author{Jesper M{\o}rk}
\affiliation{DTU Fotonik, Department of Photonics Engineering, Technical University of Denmark, Building 343, 2800 Kgs. Lyngby, Denmark}
\author{Dirk R. Englund}
\affiliation{Department of Electrical Engineering and Computer Science, Massachusetts Institute of Technology,
Cambridge, Massachusetts 02139, USA}

\date{\today}

\begin{abstract}
We investigate the origin of imperfections in the fidelity of a two-photon controlled-phase gate based on two-level-emitter non-linearities. 
We focus on a passive system that operates without external modulations to enhance its performance. 
We demonstrate that the fidelity of the gate is limited by 
opposing requirements on the input pulse width for one- and two-photon scattering events. For one-photon scattering, the 
spectral pulse width must be narrow compared to the emitter linewidth, while two-photon scattering processes require 
the pulse width and emitter linewidth to be comparable. 
We find that these opposing requirements limit the maximum fidelity of 
the two-photon controlled-phase gate to $84\%$ for photons with Gaussian spectral profiles. 
\end{abstract}

\maketitle
\section{Introduction}
\seclab{intro}

Key requirements for the successful implementation of photonic quantum 
computing architectures are i) efficient sources of single indistinguishable photons, and 
ii) a method to coherently interact two such 
photons~\cite{Knill01,Kok07,Kokbook,Ralph02,Pooley12,Gimeno-Segovia2015,OBrien03}. 
Since these requirements were first stated, single photon sources 
have steadily improved~\cite{OBrien03,Kuhn2002,Hijlkema2007,Aharonovich2016}, 
with the most promising platforms based on few-level-emitters, most notably 
semiconductor quantum dots~\cite{somaschi2016near,Ding2016,IlesSmith2016} which now boast 
near-unity indistinguishability with (source to first objective) efficiencies above $70\%$. 
Generating photon--photon interactions can be achieved by `off-line non-linearities' consisting of measurements 
and feed-forward~\cite{Knill01,Kok07,Kokbook,Ralph02,OBrien03}, or deterministically using `in-line' non-linearities based on a 
non-linear material through which two or more photons interact~\cite{Chuang1995,Duan2004,PhysRevA.93.040303,Hacker2016}. These in-line 
non-linearities can in principle also be generated by few-level-emitters~\cite{Auffeves-garnier2007,Nysteen2015PRA,Nysteen14}, 
suggesting a quantum photonic architecture in which few-level-systems act as both photon sources and photon couplers.

Experimentally, probabilistic photonic gates have been demonstrated using off-line non-linearities 
in both free-space~\cite{OBrien03,Zhao2005,PhysRevLett.93.020504} and in integrated platforms~\cite{Crespi2011,Pooley12}. 
Strong in-line non-linearities and photon switching have been achieved using Rubidium atoms strongly coupled 
to optical cavities~\cite{Reiserer14,Shomroni14,Tiecke14,Hacker2016}, 
quantum dots in photonic crystal cavities~\cite{Kim13,Volz12,Englund12,Bose12}, 
and nitrogen vacancy centers in diamond~\cite{Wang13}. 
The potentially deterministic nature of few photon in-line non-linearities makes this approach 
particularly attractive for the realisation of photonic gates, and a number of proposals have been put forward to construct 
controlled-phase gates on various platforms and with various degrees of 
complexity~\cite{Chuang1995,Duan2004,Koshino2010,Chudzicki2013,Johne12,Ralph15,Brod2016}. 
Some proposals even have the potential to operate at near unity fidelities by 
using distributed interactions~\cite{Brod2016} or pulse reshaping techniques~\cite{Ralph15}, 
though these approaches have the challenges of high complexity and potentially high losses. 
Ultimately the usefulness of a photonic gate in future quantum computing architectures will depend on the ease with which 
it can be experimentally realised and repeated, and the maximum efficiency and fidelity that it can achieve.

In this work we analyse the performance of perhaps the simplest deterministic 
passive controlled-phase gate, which acts on two uncorrelated indistinguishable photons in a dual-rail encoding. 
The idealised gate we consider uses the in-line non-linearities of 
two two-level-emitters embedded in loss-less waveguides. 
The fundamental operating 
principle of the gate relies on the saturability of a two-level-emitter, which means 
that the phase imparted onto a photon or photons scattering 
on such an emitter depends on how many photons 
are present~\cite{Fan10,Nysteen14,Nysteen2015PRA,Xu13}. 
Although it has been shown that such a gate can never perform with perfect fidelity~\cite{Shapiro2006,Xu13}, the purpose of 
this work is to understand the limits and origins of its imperfects with a view towards improved future implementations. 
Even in the loss-less case where the gate is fully deterministic, we show that 
the maximum gate fidelity is limited to $84\%$ for single photons with Gaussian spectral profiles. 
This number is determined by opposing requirements on the spectral width of the input photons; 
one-photon scattering requires spectrally narrow photons so that the greatest fraction is 
strictly resonant with the emitters, while two photon scattering requires photons with spectral 
widths similar to the emitter linewidths, which maximises saturation effects. Although the 
fidelities we calculate are significantly less than unity, in contrast to 
other schemes, the present one does not use dynamical photon capture methods~\cite{Johne12}, 
uses only two (identical) emitters per gate~\cite{Brod2016}, and does not take advantage 
of possible pulse reshaping techniques~\cite{Ralph15}, all of which are likely to introduce additional losses. 

This paper is organized as follows. In \secref{cnot_intro} the basic gate structure and 
components are introduced, and 
the gate operation in an idealised case is discussed. 
In \secsref{Lin_refs}{nLin_refs} a more realistic scenario is 
analysed and the linear and non-linear gate operations are described. A general fidelity 
measure is considered in~\secref{alt_Fid} to quantify the gate performance, and 
we conclude our findings in \secref{conclusion}.

\section{The controlled-phase gate}
\seclab{cnot_intro}

\onefigadv{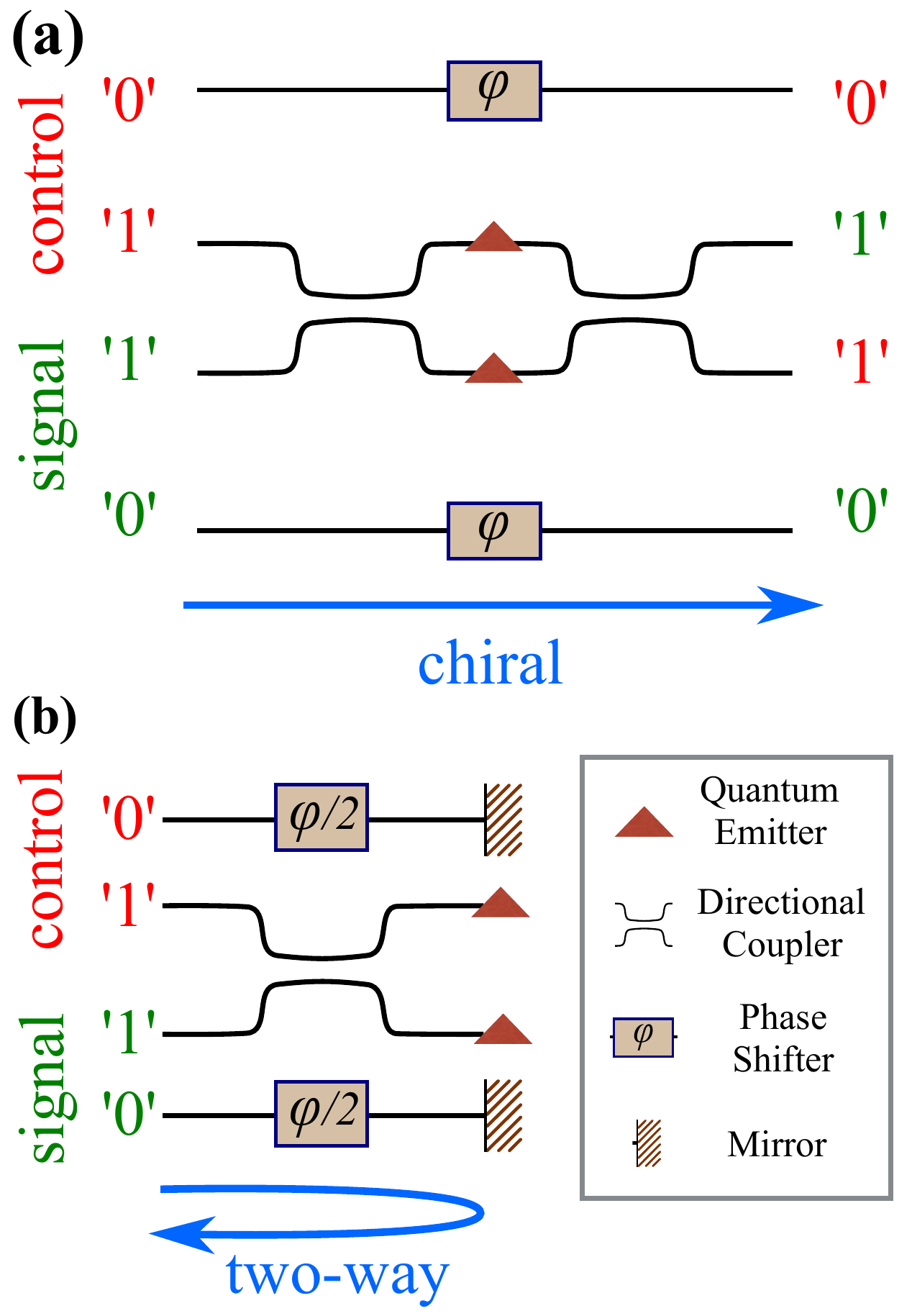}{(a) Schematic of the controlled-phase gate, 
which uses chiral waveguides, directional couplers, 
phase shifters, and two identical quantum emitters. 
The central idea of the gate is that the directional couplers act as 50/50 beam splitters, 
and as such the input state $\ket{1_\text{c}}\ket{1_\text{s}}$ gives rise to a Hong--Ou--Mandel bunching 
effect which can access the inherent non-linearities of the emitters. 
Only the $\ket{1_\text{c}}\ket{1_\text{s}}$ input state bunches in this way, while all others transform linearly, thus 
providing a fundamental non-linear interaction which can realize a two-photon gate.
We focus on the chiral setup illustrated in (a), though an equivalent scheme can be 
realized with convention bi-directional couplers as shown in (b). 
Note that the '1' arm for the control and signal is interchanged at the output 
ports in both cases.}
{cnot_waveguide_proposal}{0.45}{!h}

The structure implementing the gate, shown in \figref{cnot_waveguide_proposal}, consists of two phase shifters, two directional couplers, 
and two two-level emitters, similar to the systems in Refs.~\cite{Knill01,Ralph15,Witthaut12}. 
We focus here on two-level-emitters, though we note also that very high Q-cavities could also be used~\cite{Biberman2012,Sekoguchi2014}, 
in which splitting of the energy spectrum arises though the non-linear Kerr effect~\cite{Sun13}. 
In \figref{cnot_waveguide_proposal}(a) we envisage chiral waveguides, for which propagation is permitted only in one direction. 
We note, however, that an equivalent scheme can be realized by using 
standard bi-directional waveguides with emitters or perfectly reflecting mirrors placed at their ends, 
as illustrated in \figref{cnot_waveguide_proposal}(b). For concreteness we focus on the chiral setup of \figref{cnot_waveguide_proposal}(a), 
though all of our subsequent analysis equally applies to the two-way setup in \figref{cnot_waveguide_proposal}(b). 
The central idea behind the scheme is that the components and waveguides are arranged in such a way that only the combined 
control and signal input state $\ket{1_\text{c}}\ket{1_\text{s}}$ accesses the non-linearity of the two-level systems. 

To gain some intuition, we first consider quasi-monochromatic input photons, 
having a bandwidth much narrower than that of the emitters. 
Since the state of one photon can affect the state of the other, we must in general consider 
how pairs of photons are transformed by the gate components. 
Consider first the evolution of two photons in the state $\ket{0_\text{c}}\ket{0_\text{s}}$. 
From \figref{cnot_waveguide_proposal} we see that these 
photons each pick up a phase of $\varphi$, producing the transformation $\ket{0_\text{c}} \ket{0_\text{s}} \to\ee^{2 \ii \varphi} \ket{0_\text{c}} \ket{0_\text{s}}$.  
For input states $\ket{1_\text{c}}\ket{0_\text{s}}$ or $\ket{0_\text{c}}\ket{1_\text{s}}$, the photon 
in the $\ket{0}$ state again picks up a phase 
of $\varphi$, while the other passes through the directional couplers and a two-level emitter. 
The directional couplers act as 50/50 beam splitters, affecting the mode transformation 
\beqa
\begin{bmatrix}
       a^\dagger_{1\text{c}}  \\[0.3em]
       a^\dagger_{1\text{s}} 
\end{bmatrix} \longrightarrow
\frac{1}{\sqrt{2}}
\begin{bmatrix}
       1 & -\ii  \\[0.3em]
       -\ii & 1 
\end{bmatrix}
\begin{bmatrix}
       a^\dagger_{1\text{c}}   \\[0.3em]
       a^\dagger_{1\text{s}} 
\end{bmatrix}, \eqlab{cnot_BS}
\eeqa
where $a^\dagger_{1\text{c}}\ket{\phi}=\ket{1_\text{c}}$ and $a^\dagger_{1\text{s}}\ket{\phi}=\ket{1_\text{s}}$ with $\ket{\phi}$ denoting the vacuum. 
In this simplistic monochromatic scenario, let us assume a single photon incident on the emitter acquires a phase of $\theta$. 
Then the combined effects of the two directional couplers and the emitter cause the transformation 
$\ket{1_\text{s}} \to -\ii \ee^{\ii \theta} \ket{1_\text{c}}$ and $\ket{1_\text{c}}\to -\ii \ee^{\ii \theta} \ket{1_\text{s}}$. 
Therefore the photonic states transform as 
$\ket{1_\text{c}}\ket{0_\text{s}} \to -\ii \ee^{\ii \varphi}\ee^{\ii \theta}\ket{1_\text{s}}\ket{0_\text{s}}$ and  
$\ket{0_\text{c}}\ket{1_\text{s}} \to -\ii \ee^{\ii \varphi}\ee^{\ii \theta}\ket{0_\text{c}}\ket{1_\text{c}}$. 
Considering now the input state $\ket{1_\text{c}}\ket{1_\text{s}}$, we find that the action of the first directional 
coupler is to give rise to the Hong--Ou--Mandel interference effect;  
immediately after the first directional coupler we have a state proportional to 
$((a_{1\text{c}}^{\dagger})^2+(a_{1\text{s}}^{\dagger})^2)\ket{\phi}$, 
in which two photons are incident on each emitter in superposition. 
We denote the phase acquired by a two-photon state passing through an emitter as $\chi$, 
and therefore find that following the second directional coupler we have the transformation 
$\ket{1_\text{c}}\ket{1_\text{s}} \to (-\ii)^2\ee^{\ii \chi}\ket{1_\text{c}}\ket{1_\text{s}}$.

Collecting these results and relabelling $-\ii\ket{1_\text{s}} \to \ket{1_\text{c}}$ and $-\ii\ket{1_\text{c}} \to \ket{1_\text{s}}$ we find 
\begin{align}
\ket{0_\text{c}}\ket{0_\text{s}} &\longrightarrow \ee^{2\ii \varphi}\ket{0_\text{c}}\ket{0_\text{s}} \nn \\
\ket{0_\text{c}}\ket{1_\text{s}} &\longrightarrow \ee^{\ii \varphi}\ee^{\ii \theta}\ket{0_\text{c}}\ket{1_\text{s}} \nn \\
\ket{1_\text{c}}\ket{0_\text{s}} &\longrightarrow \ee^{\ii \varphi}\ee^{\ii \theta}\ket{1_\text{c}}\ket{0_\text{s}} \nn \\
\ket{1_\text{c}}\ket{1_\text{s}} &\longrightarrow \ee^{\ii \chi}\ket{1_\text{c}}\ket{1_\text{s}}.
\eqlab{cphase_scheme}
\end{align}
If the emitters acted as linear optical elements, we would have $\chi=2\theta$. Absorbing 
the phases $\varphi$ and $\theta$ into the definitions of $\ket{0}$ and $\ket{1}$ respectively, the transformation is locally equivalent to the identity 
and therefore does not mediate any two-photon interaction. However, if the emitter--photon interaction can be tailored  
such that $\theta = \varphi$ and $\chi=2 \varphi + \pi$, the transformation in \eqref{cphase_scheme} becomes proportional to the desired control phase gate unitary $\mathrm{diag}(1,1,1,-1)$. 
As such, if the conditions $\theta = \varphi$ and $\chi=2 \varphi + \pi$ can be met a controlled-phase gate is realized. 
Though we do not expect this to be possible with perfect accuracy~\cite{Xu13}, 
in what follows we shall explore the differing requirements on the pulse shape relative to 
the emitter linewidth which these conditions impose.

In addition to the two-level-emitters, the other essential components of the gate 
are the directional couplers needed to produce the transformation 
in \eqref{cnot_BS} and induce the Hong--Ou--Mandel effect for the input state $\ket{1_\text{c}}\ket{1_\text{s}}$. 
These components may be realized in various waveguide technologies, 
such as silica-on-silicon ridge waveguides~\cite{Thompson11}, GaAs photonic ridge waveguide 
circuits~\cite{Wang14}, photonic crystals waveguides~\cite{Martinez03}, or silicon on insulator platforms~\cite{Harris2016}, 
where in all cases the length of the coupling region must be engineered such the 
symmetrical beam splitter relation in \eqref{cnot_BS} is achieved. 
We also note, that due to the choice of directional coupler, the output port of the `1' control and signal states are swapped, 
as indicated in~\figref{cnot_waveguide_proposal}(a). This amounts to nothing more than notation, and could easily be rectified by 
introducing a crossover between the two `1' outputs. 

For proper functionality of the gate, the input states $\ket{0_\text{c}}\ket{0_\text{s}}$, $\ket{1_\text{c}}\ket{0_\text{s}}$, 
and $\ket{0_\text{c}}\ket{1_\text{s}}$, which only experience linear scattering effects, and the input state 
$\ket{1_\text{c}}\ket{1_\text{s}}$, which undergoes a non-linear transformation, must all provide the 
desired output states in~\eqref{cphase_scheme} when $\theta=\varphi$ and $\chi=2\varphi+\pi$. These scattering-induced 
changes are investigated below, treating the linear and non-linear case separately.

\section{Linear gate interactions}
\seclab{Lin_refs}

Let us now consider the gate components in more detail and analyse the conditions under which the 
scheme can be realized for more realistic non-monochromatic single photon inputs. 
We describe a single photon in the $\ket{0_\text{c}}$ state as
\beq{SinglePhotonPulse}
\ket{0_\text{c}}=\int_{-\infty}^{\infty}\mathrm{d} k\, \xi(k) \,a_{0\text{c}}^{\dagger}(k)\ket{\phi},
\eeq
where $\ket{\phi}$ is again the vacuum, 
$\xi(k)$ is the spectral profile of the photon satisfying $\int_{-\infty}^{\infty}\mathrm{d}k|\xi(k)|^2=1$, 
while $\smash{a_{0\text{c}}^{\dagger}(k)}$ is the creation operator of 
photons in the control `0' waveguide with momentum $k$, satisfying $\smash{[a_{0\text{c}}(k),a_{0\text{c}}^{\dagger}(k')]=\delta(k-k')}$. 
We note that these conditions ensure the input state $\ket{0_\text{c}}$ contains exactly one photon, 
and we consider a rotating frame such 
that $k$ is measured relative to the carrier momentum, $k_0 = \omega_0/c$. The simple transformations 
in~\eqref{cphase_scheme} are not generally valid for photonic wavepackets comprised by many $k$-modes 
because the phases $\varphi$ and $\theta$ depend on $k$. In a large-scale system, the output from one 
gate must function as the input to another gate and they should therefore only differ by a time-translation, 
which in momentum space corresponds to the transformation $\xi(k)\to \xi(k)\mathrm{e}^{i \varphi(k)}$ with 
\beq{AllowedTransformation}
\varphi(k) = \varphi_0 + kL, 
\eeq
where $L$ is an additional optical path length of the `0' waveguides, 
either induced by a change in the refractive index of the material or by a longer arm length. 

When $\varphi(k)$ is of the form in~\eqref{AllowedTransformation}, the input state $\ket{0_\text{c}}\ket{0_\text{s}}$ 
is described by a product of two single-photon states 
of the form in \eqref{SinglePhotonPulse}, and we write the corresponding output state as 
$\ket{0_\text{c}}\ket{0_\text{s}}\to \ket{\tilde{0}_\text{c}}\ket{\tilde{0}_\text{s}}$ with
\beq{SinglePhotonPulseScattered}
\ket{\tilde{0}_\text{c}}= - \int_{-\infty}^{\infty}\mathrm{d} k\, \xi(k)\ee^{i k L} \,a_{0\text{c}}^{\dagger}(k)\ket{\phi},
\eeq
and a similar definition for $\ket{\tilde{0}_\text{s}}$. 
Single photons with states of this form will be considered our `ideal' output states, 
since they are identical to the input state up to a linear frequency-dependent phase 
corresponding to a fixed temporal delay. 
The choice of $\varphi_0 = \pi$ has been chosen in anticipation of the 
transformation of the $\ket{0_\text{c}}\ket{1_\text{s}}$ state discussed below.

We now consider changes to the two input states with a single photon in one of the `1' arms, 
$\ket{0_\text{c}}\ket{1_\text{s}}$ and $\ket{1_\text{c}}\ket{0_\text{s}}$. The photon 
in the `0' arm is treated analogously to \eqref{SinglePhotonPulseScattered}, while that 
in the `1' arm instead interacts with an emitter. Photons passing through the `1' arms must also 
give rise to states differing from input states only by a time-translation. To see the conditions under 
which this is the case, we consider a non-monochromatic single photon as described by~\eqref{SinglePhotonPulse} 
scattering on a two-level emitter in a chiral waveguide. 
The photon will acquire a complex coefficient $t(k)$ for 
each momentum component $k$, resulting in a photon  
with spectral profile $t(k)\xi(k)$. The frequency-dependent 
transmission coefficient is~\cite{Shen07,Rephaeli13},
\beqa
t(k)=\frac{k-\Delta-\ii(\Gamma-\gamma)/v_\text{g}}{k-\Delta+\ii(\Gamma+\gamma)/v_\text{g}},
\eqlab{cnot_tk_real}
\eeqa
where $v_\text{g}$ is the group velocity in the waveguide, $\Delta$ the momentum detuning of the 
emitter from the pulse carrier frequency, $\Gamma$ the emitter decay rate into waveguide modes, and $\gamma$ the loss rate into modes outside the waveguide~\cite{Rephaeli13}.
\onefigadv{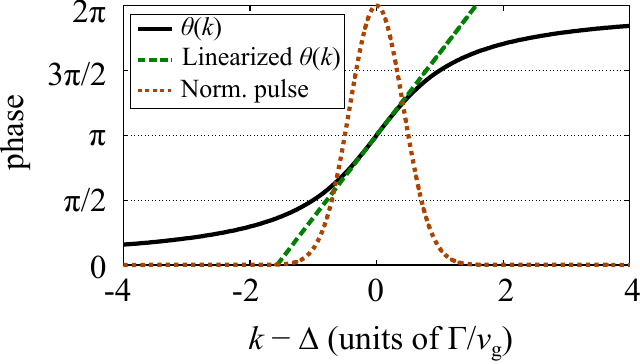}
{Phase $\theta(k)$ acquired by a single-photon 
wavepacket component with moment $k$ propagating in a chiral waveguide scattering on a lossless
resonant emitter (black solid line),  
together with a linear approximation, see \eqref{cnot_tk_approx} (green dashed line). 
By comparison, the spectrum of a resonant Gaussian wavepacket with spectral FWHM of $\sigma=\Gamma/v_\text{g}$ 
is shown (orange dotted line) with a scaled intensity to match the plotting window.}
{cnot_bs_and_singlePhase}{0.47}{t}
Recalling the effect of the directional couplers, we find that the 
states transform as 
$\ket{0_\text{c}}\ket{1_\text{s}}\to -i \ket{\tilde{0}_\text{c}}\ket{\bar{1}_\text{c}}$ and 
$\ket{1_\text{c}}\ket{0_\text{s}}\to -i \ket{\bar{1}_\text{s}}\ket{\tilde{0}_\text{s}}$, where 
\beq{SinglePhotonPulseScatteredQD}
\ket{\bar{1}_\text{c}}=\int_{-\infty}^{\infty}\mathrm{d} k\, \xi(k)t(k) \,a_{1\text{c}}^{\dagger}(k)\ket{\phi},
\eeq
with a similar definition for $\ket{\bar{1}_s}$. In the loss-less case $\gamma=0$ and 
$|t(k)|=1$, meaning that $\braket{\bar{1}_\text{c}}{\bar{1}_\text{c}}=1$ and the output state contains exactly one photon. 
As previously discussed, 
we can simply relabel what we refer to as the control and signal photons in the outputs, 
and absorb factors of $-\ii$ in these definitions. We then have 
$\ket{0_\text{c}}\ket{1_\text{s}}\to \ket{\tilde{0}_\text{c}}\ket{\bar{1}_\text{s}}$ and 
$\ket{1_\text{c}}\ket{0_\text{s}}\to \ket{\bar{1}_\text{c}}\ket{\tilde{0}_\text{s}}$. 

What is required, however, is that 
each photon has a spectral profile identical to an `ideal' state,  $\ket{\tilde{1}_\text{c}}$ or $\ket{\tilde{1}_\text{s}}$, 
defined as in \eqref{SinglePhotonPulseScattered} with $\smash{a_{0\text{c}}^{\dagger}}$ replaced with $\smash{a_{1\text{c}}^{\dagger}}$ or $\smash{a_{1\text{s}}^{\dagger}}$. 
Considering again the loss-less case where $\gamma=0$ we can write $t(k)=\exp[\ii\theta(k)]$. 
The phase $\theta(k)$ is shown as a function of $k$ in \figref{cnot_bs_and_singlePhase}. 
If the incoming single-photon has a carrier frequency corresponding to the emitter transition frequency, 
$\Delta=0$, the phase can be Taylor expanded around $k/\tilde{\Gamma}=0$, producing 
\beqa
\theta(k)=\pi+ 2\frac{k}{\tilde{\Gamma}} +\mathcal{O}\left(\frac{ k}{\tilde{\Gamma}}\right)^{\!\!3} \eqlab{cnot_tk_approx} ,
\eeqa
where $\tilde{\Gamma} = \Gamma/v_{\text{g}}$. Keeping the condition $\varphi = \theta$ in mind and 
comparing~\eqsref{AllowedTransformation}{cnot_tk_approx}, we see that a good gate performance 
requires $|k|\ll \tilde{\Gamma}$, which corresponds to pulses with a spectrum that is much narrower than the emitter linewidth. 
\onefigadv{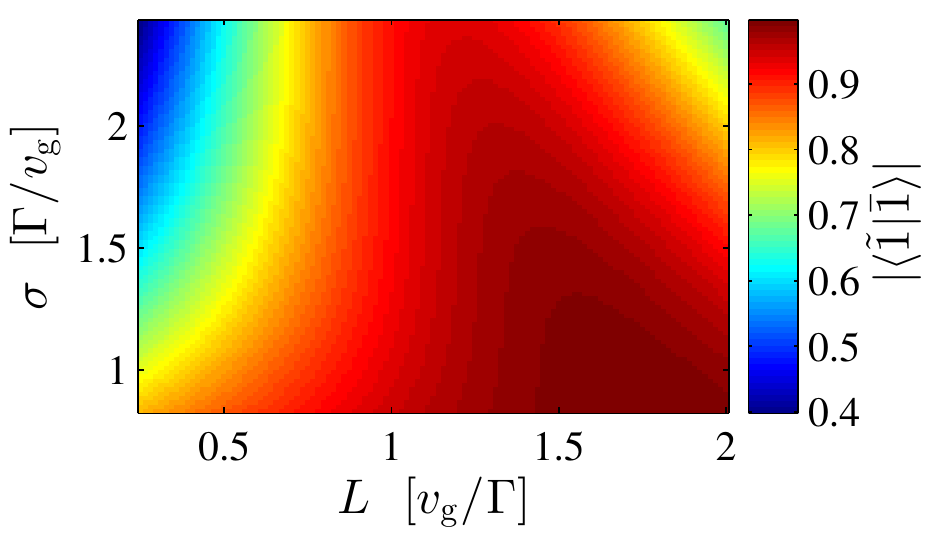}{Overlap between the ideal and scattered state for 
logical inputs $\ket{1_\text{c}}\ket{0_\text{s}}$ or $\ket{0_\text{c}}\ket{1_\text{s}}$ 
as a function of the additional optical length of the `0' arms, $L$, and the input spectral width $\sigma$ 
defined in Eq.~({\ref{GaussianProfile}}).}
{F10_vs_sigma_and_L}{0.45}{b}
For spectrally broader photons for which $\xi(k)$ extends 
beyond $k\sim\tilde{\Gamma}$, 
terms of higher order in $k$ will have an influence and introduce chirping effects~\cite{Nysteen14,Nysteen2015PRA}. 

To illustrate this in more detail for a specific case, let us consider 
a Gaussian single-photon wavepacket, 
defined by the spectral profile 
\beqa
\label{GaussianProfile}
\xi(k)=(\pi\sigma'^2)^{-1/4}\exp[-k^2/(2\sigma'^2)],
\eeqa
where the spectral bandwidth (FWHM of the intensity spectrum) is $\smash{\sigma=2\sqrt{\ln(2)}\sigma'}$. 
\figref{F10_vs_sigma_and_L} plots the magnitude of the overlap between the 
desired (ideal) state and actual state for a Gaussian spectrum as described above. 
As expected, the overlap increases when the spectral width, $\sigma$, is decreased. 
The optimum additional path length, $L$, approaches 
$L\!=\! 2v_{\text{g}}/\Gamma$ as $\sigma$ is decreased, which is expected from the linear 
term in~\eqref{cnot_tk_approx}. For larger spectral widths, the optimum $L$ decreases because 
a straight line with a slope smaller than $2v_{\text{g}}/\Gamma$ approximates the phase 
$\theta(k)$ better in this case, as seen in~\figref{cnot_bs_and_singlePhase}.

\section{Non-linear gate interactions}
\seclab{nLin_refs}

The non-linear interaction occurs for the input state $\ket{1_\text{c}}\ket{1_\text{s}}$, where two 
photons may be present at the scatterers simultaneously, introducing non-linear interactions 
through a two-photon bound state \cite{Fan10}. The non-linear scattering is treated by the scattering 
matrix formalism following Ref. \cite{Fan10}, and we include the directional coupler when 
calculating the scattered state of the entire gate. The gate input consists of two uncorrelated identical 
photons which we describe by 
\begin{align}
\ket{\psi_\text{in}} &= \int\limits_{-\infty}^{\infty} \int\limits_{-\infty}^{\infty}\textrm{d}k\,\textrm{d}k'\,\xi(k)\xi(k')a^\dagger_{\text{c}1}(k)a^\dagger_{\text{s}1}(k')\ket{\phi},
\end{align}
where as before $\int_{-\infty}^{\infty}\mathrm{d}k |\xi(k)|^2=1$ to ensure $\ket{\psi_\text{in}}$ contains two photons. 
Following the action of the first directional coupler, scattering on the two-level emitters, and 
passing through the second directional coupler, we find $\ket{\psi_\text{in}}\to\ket{\psi_\text{scat}}$ 
with the total scattered state given by
\begin{align}
\ket{\psi_\text{scat}} &=\int\limits_{-\infty}^{\infty} \int\limits_{-\infty}^{\infty}\textrm{d}k\,\textrm{d}k'\, \beta_\text{scat}(k,k')a^\dagger_{\text{c}1}(k)a^\dagger_{\text{s}1}(k')\ket{\phi}, \eqlab{cnot_2phot_scat}
\end{align}
where we have removed a factor of $(-\ii)^2$ to be consistent with our definitions of the output states, and
\beqa
\beta_\text{scat}(k,k')=\beta_\text{scat}^\text{linear}(k,k')+\frac{1}{2}b(k,k'), \eqlab{two_fan_wp1}
\eeqa
with the linear contribution given by $\beta_\text{scat}^\text{linear}(k,k')=t(k)t(k')\xi(k)\xi(k')$ 
and a non-linear scattering contribution by
\beqa
b(k,k')= \!\!\int_{-\infty}^{\infty}\!\!\! \text{d}p \; \xi(p)\xi(k\!+\!k'\!\!-\!p)B_{kk'p(k\!+\!k'\!-\!p)}.
\eeqa
The scatterer-dependent coefficient $B_{kk'pp'}$ is evaluated in Ref.~\cite{Rephaeli13} for a two-level system, 
\beqa
B_{kk'pp'}=\ii\frac{\sqrt{2\Gamma/v_\text{g}}}{\pi}s(k)s(k')[s(p)+s(p')], \eqlab{two_fan_theB}
\eeqa
where
\beqa
s(k) = \frac{\sqrt{2\Gamma/v_\text{g}}}{k-\Delta+\ii(\Gamma+\gamma)/(v_\text{g})}.
\eeqa
The ideal output state in the non-linear case is 
\onefigadv{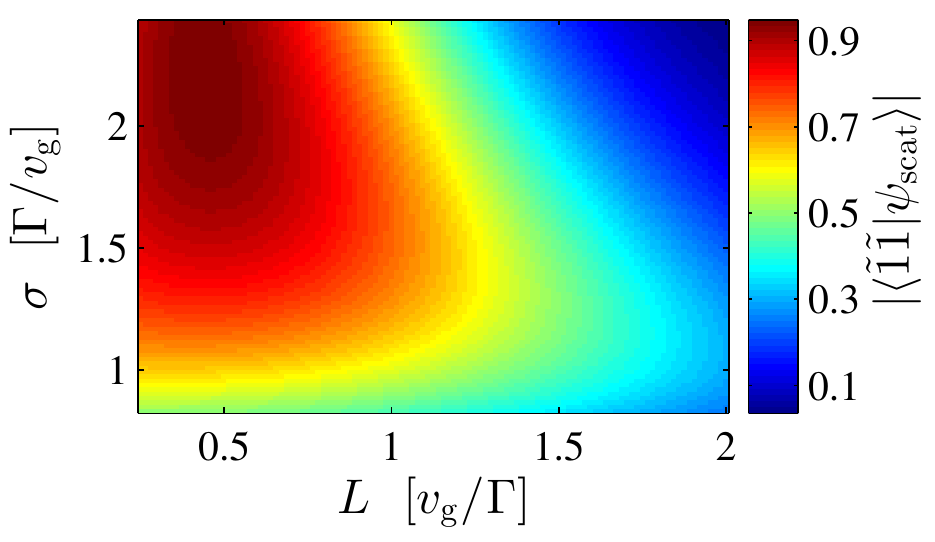}{Overlap between the ideal and scattered state for 
the $\ket{1_\text{c}}\ket{1_\text{s}}$ input as a function of the additional optical length of the `0' arms, 
$L$, and the input pulse width, $\sigma$.}
{F11_vs_sigma_and_L}{0.45}{b}
\beq{TwoPhotonPulseScattered}
\ket{\tilde{1}_\text{c}}\ket{\tilde{1}_\text{s}}= - \!\! \int\limits_{-\infty}^{\infty} \int\limits_{-\infty}^{\infty} \!\! \textrm{d}k\,  \textrm{d}k'\, \ee^{i (k+k') L}  a^\dagger_{\text{c}1}(k)a^\dagger_{\text{s}1}(k')\ket{\phi},
\eeq
where the minus sign accounts for the required phase flip that defines the controlled-phase gate. 

To gain some insight into how well the 
actual state, $\ket{\psi_{\rm{scat}}}$, approximates the ideal state in~\eqref{TwoPhotonPulseScattered}, 
we plot the magnitude of their overlap as a function of $L$ and $\sigma$ in~\figref{F11_vs_sigma_and_L}, 
again for Gaussian input pulses. 
In contrast to the one-photon scattering case in~\figref{F10_vs_sigma_and_L}, we now see that 
the largest overlap is observed for pulse widths $\sigma \approx 2.2 \Gamma/v_{\text{g}}$. This is because it is 
for these widths that the non-linearities are strongest and the required $\pi$-phase shift can be generated, 
consistent with the results in Ref.~\cite{Nysteen14}\footnote{In Ref. \cite{Nysteen14}, the interaction is reported 
strongest when $\sigma\sim \Gamma/v_{\text{g}}$. This occurs for an emitter in a bi-directional waveguide, 
which effectively has twice the larger decay rate as for the single-directional problem considered in this work. 
Thus, when projected to this work, the strongest non-linearity is expected around $\sigma\sim 2 \Gamma/v_{\text{g}}$ }. 
Furthermore, the optimal value of $L$ in this non-linear scattering case is significantly lower than in the linear case. 
A comparison of~\figsref{F10_vs_sigma_and_L}{F11_vs_sigma_and_L} demonstrates that 
limitations in the gate performance are expected to occur 
because of these different requirements on $\sigma$ and $L$ to 
optimally approximate the ideal output states in the linear and non-linear cases, 
which we now explore in more detail.

\section{Fidelity of the Gate Operation}
\seclab{alt_Fid}

\onefigadv{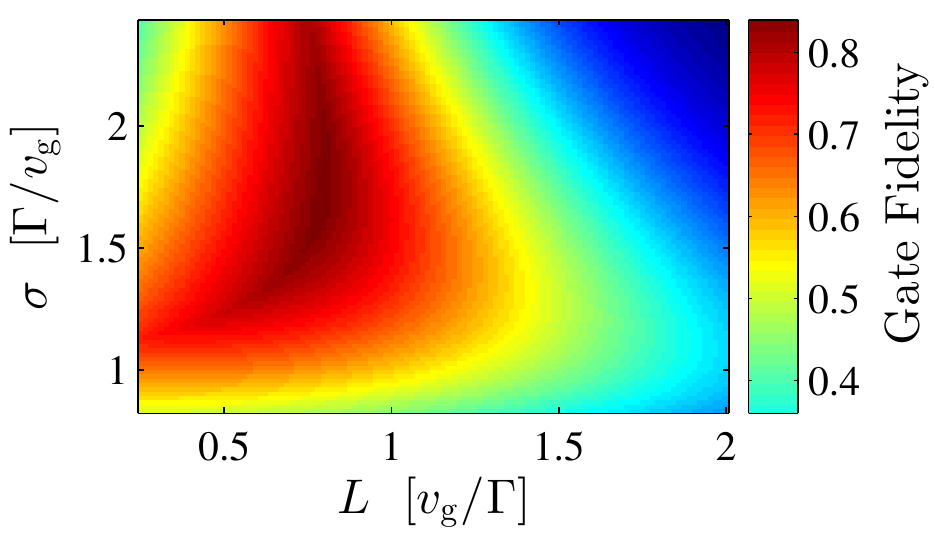}{Gate fidelity as a function of the additional optical length of the `0' arms, $L$, and the input pulse width, $\sigma$.}
{fidelity_vs_sigma_and_L}{0.45}{t}

In order to find the optimal spectral width $\sigma$ and path length difference $L$, we now consider 
the operation of the gate as a whole. 
When incorporated into a larger optical circuit, the logical input state of the gate will necessarily be unknown, 
and the gate must therefore be able to operate for any linear combination of the four possible logical input states.
As such, the gate performance must be quantified by a fidelity based on a worst case scenario, in which 
the output state of the gate is compared to the ideal target output state, minimised over all possible input states. 
A gate fidelity meeting these requirements is defined as~\cite{Chuang}
\beq{mik1}
F(\hat{U}, \mcE) \equiv \min_{\ket{\psi}} F_{\!s}\big(\hat{U}\ket{\Psi}\bra{\Psi}\hat{U}^{\dagger}, \mcE(\ket{\Psi}\bra{\Psi})  \big),
\eeq
where $\hat{U}$ and $\mcE$ describe the transformations of the ideal and actual gate, respectively, and $F_{\!s}$ is the state fidelity defined by~\cite{Chuang}
\beq{state F}
F_{\!s}(\hat{\rho}, \hat{\sigma}) \equiv \text{Tr} \left\{\sqrt{ \hat{\rho}^{\frac12} \hat{\sigma}\hat{\rho}^{\frac12}}\right\},
\eeq
for two density operators, $\hat{\rho}$ and $\hat{\sigma}$. The arbitrary input state, $\ket{\Psi}$, is given by 
\begin{align} \eqlab{mik2}
\ket{\Psi}  &= \big(\alpha \ket{0_\text{s}} +\beta \ket{1_\text{s}} \big)\otimes \big(\zeta \ket{0_\text{c}} +\vartheta \ket{1_\text{c}} \big)  \nn \\
 &=\alpha\zeta \ket{00} + \alpha\vartheta\ket{01} + \beta\zeta \ket{10} + \beta\vartheta \ket{11}, 
\end{align}
where $\ket{0_\text{s}}\ket{0_\text{c}}\equiv \ket{00}$ etc. 
Using the definitions from previous sections, the ideal gate transformation is
\begin{align} \eqlab{mik ideal}
\hat{U}\ket{\Psi}  &= \alpha\zeta \ket{\tilde{0}\tilde{0}} + \alpha\vartheta\ket{\tilde{0}\tilde{1}} + \beta\zeta \ket{\tilde{1}\tilde{0}} - \beta\vartheta \ket{\tilde{1}\tilde{1}}.
\end{align}
If we neglect loss, the output states are pure and the actual (possibly imperfect) transformation is described 
by $\mcE (\ket{\Psi}\bra{\Psi})= \hat{T}\ket{\Psi}\bra{\Psi} \hat{T}^{\dagger}$, with 
\begin{align} \eqlab{mik actual} 
\hat{T}\ket{\Psi}  &= \alpha\zeta \ket{\tilde{0}\tilde{0}} + \alpha\vartheta\ket{\tilde{0}\bar{1}} + \beta\zeta \ket{\bar{1}\tilde{0}} + \beta\vartheta \ket{\psi_\text{scat}},
\end{align}
where $\ket{\psi_\text{scat}}$ is given by \eqref{cnot_2phot_scat}. 
For pure states,~\eqref{state F} simplifies to $F_{\!s}(\ket{a}\bra{a}, \ket{b}\bra{b}) = |\braket{a}{b}|$, 
and the state fidelity is therefore
\begin{multline} \eqlab{mik3a}
F_{\!s}(\hat{U}\ket{\Psi}\bra{\Psi}\hat{U}^{\dagger}, \hat{T}\ket{\Psi}\bra{\Psi}\hat{T}^{\dagger})  =  | \bra{\Psi} \hat{U}^{\dagger} \hat{T} \ket{\Psi}| = \\
 \Big| |\alpha\zeta |^2 \!+\! \braket{\tilde{1}}{\bar{1}} \left (|\alpha\vartheta|^2 \!+\! |\beta\zeta|^2 \right) \!-\! |\beta\vartheta|^2 \braket{\tilde{1}\tilde{1}}{\psiscat} \Big|.
\end{multline}
To find the fidelity of the gate for a given pulse width and path length difference, this state fidelity must be minimised 
over all possible logical input states $\ket{\Psi}$ parameterised by the coefficients $\alpha, \beta, \zeta, \vartheta$. 
Since the state fidelity only depends on the magnitude of the coefficients 
and the signal and control input states both must be normalized, 
the minimization in~\eqref{mik1} can be carried out by varying only, e.g. $|\alpha|$ and $|\zeta|$. 
By performing this minimization for different values of $\sigma$ and $L$, the trade-offs due to the effects of linear and non-linear scattering can be quantified. 
The result is shown in~\figref{fidelity_vs_sigma_and_L}, where the gate fidelity is plotted as a function of $L$ and $\sigma$, again 
for Gaussian pulses. 
The optimum set of parameters is seen to be close to that in~\figref{F11_vs_sigma_and_L} but shifted towards smaller pulse widths and larger $L$, where the optimum was observed in~\figref{F10_vs_sigma_and_L}. This trend is expected, since~\eqref{mik3a} effectively performs a 
weighted average of the overlaps in~\figsref{F10_vs_sigma_and_L}{F11_vs_sigma_and_L}.
\onefigadv{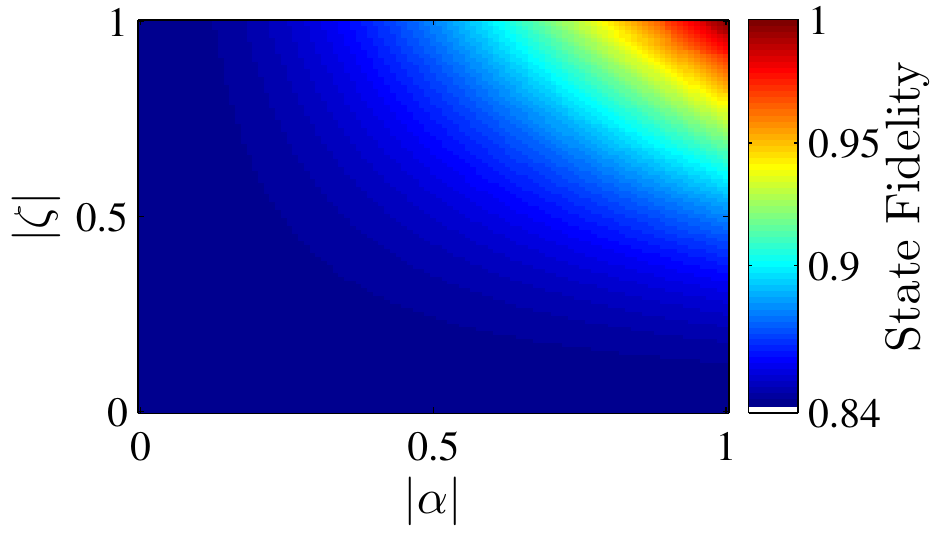}{State fidelity as a function of input states expressed by $|\alpha|$ and $|\zeta|$ for $L = 0.80 v_{\text{g}}/\Gamma$ and $\sigma=1.72\Gamma/v_{\text{g}}$ corresponding to the maximum gate fidelity in~\figref{fidelity_vs_sigma_and_L}.}
{fidelity_vs_alpha_and_zeta}{0.45}{t}
In order to confirm that the gate fidelity indeed corresponds to a worse case 
scenario,~\figref{fidelity_vs_alpha_and_zeta} shows the dependence of the state 
fidelity on the input states for the optimum parameter set in~\figref{fidelity_vs_sigma_and_L}. 
It shows that the state fidelity approaches unity for the state $\ket{0_\text{c}}\ket{0_\text{s}}$, and is 
above 84\% for the entire state space, as expected.

Finally, we note that our formalism easily allows for spectra other than Gaussians to 
be considered. Most notably, we find that Lorentzian spectral profiles result in a worse gate fidelity 
of $F\approx 62\%$. Although a Lorentzian shaped single-photon is expected to most efficiently 
populate a two-level-emitter, two such coincident pulses give rise to a smaller induced non-linearity~\cite{Nysteen2015PRA}, 
which is an essential requirement for the gate to operate. We find that $\mathrm{sech}^2$ pulses achieve a fidelity 
marginally better than Gaussian pulses, raising the gate fidelity by only $0.5\%$. Ultimately active modification of 
spectra may be necessary if gates based on two-level-emitter non-linearities are to attain fidelities 
approaching unity~\cite{Ralph15}.


\section{Conclusion}
\seclab{conclusion}

We have investigated in detail the feasibility of using two-level-emitter non-linearities to 
construct a passive two-photon controlled phase gate, elucidating 
the non-linearity-induced changes in the spectrum. 
We find that these effects ultimately limit the fidelity of a controlled phase gate based 
on two-level-emitter non-linearities, giving $F\approx 84\%$ for Gaussian input pulses, 
decreasing to $F\approx 62\%$ for Lorentzian spectra.  We emphasise, however, 
that the scheme we consider requires no dynamical capture of photons~\cite{Johne12}, uses 
only two identical two-level-emitters, and does not make use of pulse reshaping techniques. 
Although schemes making use of multiple non-linearities per gate~\cite{Brod2016}, or 
gradient echo memory~\cite{Ralph15} to reverse pulse shapes, predict theoretical fidelities 
approaching unity, these processes increase the complexity of the gate, and are likely to introduce 
additional losses. Ultimately it seems likely that efficiency--fidelity trade-offs will be present 
in any gate scheme, and these trade-offs must be carefully considered in a larger 
photonic network with a given application.

\begin{acknowledgements}
AN and JM acknowledge support from NATEC and SIQUTE. DPSM acknowledges support 
from a Marie Sk{\l}odowska-Curie Individual Fellowship (ESPCSS). MH acknowledges support from DFF:1325-00144. 
DE acknowledges support from AFOSR MURI for Optimal Measurements for 
Scalable Quantum Technologies (FA9550-14-1-0052) and the Air Force Research Laboratory RITA program (FA8750-14-2-0120). 
 
\end{acknowledgements}

\bibliographystyle{apsrev}

\end{document}